\documentclass{sig-alternate-2013}

\usepackage{mathptmx}
\usepackage{graphicx}
\usepackage{url}
\usepackage{booktabs}
\usepackage{subfigure}
\usepackage{color}
\usepackage{algorithm}
\usepackage{algpseudocode}
\usepackage{adjustbox}

\newcounter{MANumberOfComments}
\stepcounter{MANumberOfComments}
\newcounter{LINumberOfComments}
\stepcounter{LINumberOfComments}

\begin{document}

\title{Bandwidth-aware Service Placement in \\Community Network Clouds}
%\subtitle{[Extended Abstract]
%\titlenote{A full version of this paper is available as
%\textit{Author's Guide to Preparing ACM SIG Proceedings Using
%\LaTeX$2_\epsilon$\ and BibTeX} at

\numberofauthors{6} %  in this sample file, there are a *total*
\author{
% 1st. author
\alignauthor
Mennan Selimi\\
%       \affaddr{Fundaci\'{o} Privada per la Xarxa Lliure, Oberta i Neutral Guifi.net}\\
       \affaddr{UPC}\\
%       \affaddr{Mas l'Esperan\c{c}a, 08503 Gurb}\\
       \affaddr{Barcelona, Spain}\\
       %\email{mselimi@ac.upc.edu}
% 2nd. author
\alignauthor
Lloren\c{c} Cerd\`{a}-Alabern\\
%       \affaddr{Department of Computer Architecture}\\
%       \affaddr{Universitat Politecnica de Catalunya - BarcelonaTech}\\
%       \affaddr{Universitat Polit\`{e}cnica de Catalunya}\\
        \affaddr{UPC}\\
       \affaddr{Barcelona, Spain}\\
       %\email{llorenc@ac.upc.edu}
% 3rd. author
\alignauthor 
Liang Wang\\
       \affaddr{Cambridge University}\\
       \affaddr{Cambridge, UK}\\
%       \affaddr{Catalonia}\\
       %\email{first.last@cl.cam.ac.uk}
\and  % use '\and' if you need 'another row' of author names
% 4th. author
\alignauthor
Arjuna Sathiaseelan\\
%       \affaddr{Department of Computer Architecture}\\
%       \affaddr{Universitat Politecnica de Catalunya - BarcelonaTech}\\
%       \affaddr{Universitat Polit\`{e}cnica de Catalunya}\\
       \affaddr{Cambridge University}\\
       \affaddr{Cambridge, UK}\\
       %\email{first.last@cl.cam.ac.uk}
% 5th. author
\alignauthor 
Lu\'{i}s Veiga\\
       \affaddr{INESC-ID Lisboa / IST}\\
       \affaddr{Lisbon, Portugal}\\
%       \affaddr{Catalonia}\\
       %\email{first.last@inesc-id.pt}
% 6th. author
\alignauthor 
Felix Freitag\\
%       \affaddr{Fundaci\'{o} Privada per la Xarxa Lliure, Oberta i Neutral Guifi.net}\\
       \affaddr{UPC}\\
%      \affaddr{Mas l'Esperan\c{c}a, 08503 Gurb}\\
       \affaddr{Barcelona, Spain}\\
       %\email{felix@ac.upc.edu}
}

%\date{30 September 2015}

\maketitle
\begin{abstract}
Seamless computing and service sharing in community networks have been gaining momentum due to the emerging technology of community network micro-clouds (CNMCs). However, running services in CNMCs can face enormous challenges such as the dynamic nature of micro-clouds, limited capacity of nodes and links, asymmetric quality of wireless links for services, deployment models based on geographic singularities rather than network QoS, and etc. CNMCs have been increasingly used by network-intensive services that exchange significant amounts of data between the nodes on which they run, therefore the performance heavily relies on the available bandwidth resource in a network. This paper proposes a novel bandwidth-aware service placement algorithm which outperforms the current random placement adopted by Guifi.net. Our preliminary results show that the
proposed algorithm consistently outperforms the current random placement adopted in Guifi.net by 35\% regarding its bandwidth gain. More importantly, as the number of services increases, the gain tends to increase accordingly.
%%%\liang{give some numbers to showoff BASP. We can probably remove emails in case running out of space?} \mennan{we need to work on abstract}

%We present a service placement algorithm that minimizes the service overlay diameter while fulfilling service specific type criterias.
\end{abstract}

%% A category with the (minimum) three required fields
%\category{C.2}{Computer-Communication Networks}{Community Networks}
%% A category including the fourth, optional field follows...
%\category{J.m}{Computer Applications}{Miscellaneous}[Community Clouds]
%\terms{Community networks, community clouds, sustainability}

 % NOT required for Proceedings

\label{sec:intro}
\section{Introduction}

\begin{figure}
\centering
\includegraphics[width=1.0\linewidth]{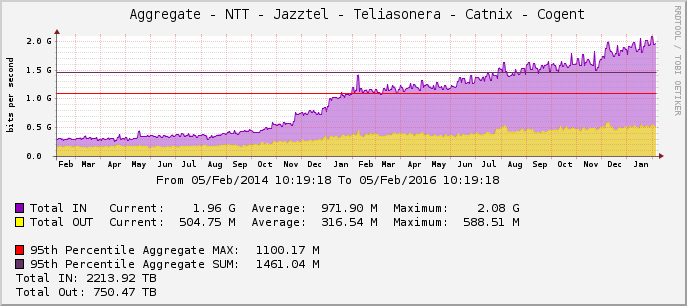}
\caption{Guifi Traffic}
\label{fig:traffic_guifi}
\end{figure}

Community networks or Do-It-Yourself networks (DIYs) are bottom-up built decentralized networks, deployed and maintained by their own users. In the early 2000s, community networks (CNs) gained momentum in response to the limited options for network connectivity in rural and urban communities. One successful effort of such a network is Guifi.net\footnote{http://guifi.net/}, located in the Catalonia region of Spain. Guifi.net is defined as an open, free and neutral community network built by its members: citizens and organizations pooling their resources and coordinating efforts to build and operate a local network infrastructure. Guifi.net was launched in 2004 and till today it has grown into a network of more than 30.000 operational nodes, which makes it the largest community network worldwide \cite{Braem2013}. Figure \ref{fig:traffic_guifi} shows the evolution of total inbound and outbound Guifi.net traffic to the Internet for the last two years. Pink represents incoming traffic from Internet and yellow represents outgoing traffic. For two years the traffic is doubled and peaks are as a result of a new links and fiber optics in the backbone.  

Similar to other community networks, Guifi.net aims to create a highly localized digital ecosystem. However, the predominant usage we have observed, is to access the cloud-based Internet services external to a community network. For instance, more than 50\% of the user-oriented services consumed in Guifi.net are gateway proxies that provide Internet connectivity hence impose a heavy burden on the limit backbone links \cite{GuifiServices}. For a very long time in the past, user-oriented services had not been developed locally because of lacking streamlined mechanisms to exploit all the available resources within a community network as well as other technological barriers.
With the adoption of \emph{community network micro-clouds}\footnote{http://cloudy.community/}, i.e., the platform that enables cloud-based services in community networks, local user-oriented services gathered a huge momentum. Community network users started creating their own homegrown services and using alternative open source software for many of today's Internet cloud services, e.g., data storage services, interactive applications such as Voice-over-IP (VoIP), video streaming, P2P-TV, and etc. In fact, a significant amount of services were already locally deployed and running within Guifi.net including GuifiTV, Graph servers, mail servers, game servers \cite{COMNET}. All these services are provided by individuals, social groups, small non-profit or commercial service providers.
%%% \liang{give a couple of sentences and a ref to argue the advantages of running local service.} \mennan{already re-wrote the paragraphs}

Because Guifi.net nodes are geographically distributed, given this set of local services, we need to decide where these services should be placed in a network. Obviously, without taking into account the underlying network resources, a service may suffer from poor performance, e.g, by sending up large amounts of data across slow wireless links while faster and more reliable links remain underutilized.
Therefore, the key challenge in community network micro-clouds is to determine the location, i.e. servers at certain geographic points in the network, where the different services multiplexed on a shared infrastructure will be running. While conceptually straightforward, it is challenging to calculate an optimal decision due to the dynamic nature of community networks and usage patterns.
In this work we aim to address the following question: \textit{"Given a community network cloud infrastructure, what is an effective and low-complexity service placement solution that maximises end-to-end performance (e.g., bandwidth)?"} %%% Our hypothesis is that by measuring the inter-node rates and bottlenecks in community network clouds, it is possible to improve the performance of a mix of applications.
%
%However taking into account the dynamic and challenging environment of community networks our algorithm (system) must overcome three challenges: first, inter-node rates in community clouds are not constant and not easy to measure; second any practical measurement or profiling method must not introduce much extra traffic since we are dealing with limited capacity links and nodes; third placing a subset of services changes the network rates available for subsequent services. Moreover, finding an optimal placement method in a realistic setting such as Guifi.net, given the network rates and application profile is computationally intractable, so any practical approach can be approximate. 
%
Our preliminary results show that the
proposed algorithm consistently outperforms the current random placement adopted in Guifi.net by 35\% regarding its bandwidth gain. More importantly, as the number of services increases, the gain tends to increase accordingly.

%Fix the section numbering M.
%%% The rest of the paper is organized as follows. In Section \ref{sec:qmp} we describe and characterize the performance of QMP network. Section \ref{sec:model} defines our system model considering the allocation model that we use. In Section \ref{sec:algorithm} and Section \ref{sec:evaluation} \liang{section ref num needs to be fixed.} we present our bandwidth-aware service placement algorithm and discuss the evaluation results. Section \ref{sec:related-work} describes related work and section \ref{sec:conclusion} concludes and discusses future research directions. 

%%%\liang{imo, intro needs to be condensed. } \mennan{I think so.}

\label{sec:qmp}
\section{Need for Localized Services}

%%% \liang{A suggestion: would it be better swap section 2 and 3. So we first present the measurements on Guifi.net to motivate our design, then based on the observation derived from measurement, we build Service Deployment Model, then further propose algorithm? Would that look better?} \mennan{I totally agree. It is done !}

In this section, we characterize wireless community networks by presenting our experimental measurements in a production example over five months, which exposes the necessity of deploying localized services \cite{ICN} and justifies our motivation of proposing an intelligent placement algorithm.

\subsection{QMP Network: A Brief Background}

The network we consider, began deployment in 2009 in a quarter of the city of Barcelona, Spain, called Sants, as part of the \textit{Quick Mesh Project\footnote{http://qmp.cat/Home}} (QMP). 
In 2012, nodes from \textit{Universitat Polit\`{e}cnica de Catalunya} (UPC) joined  the network, supported by the EU CONFINE \footnote{https://confine-project.eu/} project. 
We shall refer to this network as \textit{QMPSU} (from Quick Mesh Project at Sants-UPC). QMPSU is part of the Guifi community network which has more than $30.000$ operational nodes. At the time of writing, QMPSU has around 61 nodes, 16 at UPC and 45 at Sants. There are two gateways, one in UPC Campus and another in Sants, that connect QMPSU to the rest of Guifi.net (see Figure \ref{fig:qmpsantsupc}). A detailed description of QMPSU can be found in \cite{LlorencMSWiM}, and a live monitoring page updated hourly is available in the Internet \footnote{http://dsg.ac.upc.edu/qmpsu/index.php}.

Typically, QMPSU users have an outdoor router (OR) with a Wi-fi interface on the roof, connected through Ethernet to an indoor AP (access point) as a premises network. The most common OR in QMPSU is the NanoStation M5, which integrates a sectorial antenna with a router furnished with a
wireless 802.11an interface. Some strategic locations have several NanoStations, that provide larger coverage. In addition, some links of several kilometers are set up with parabolic antennas (NanoBridges). ORs in QMPSU are flashed with the Linux distribution which was developed inside the QMP project wihich is a branch of OpenWRT\footnote{https://openwrt.org/} and uses BMX6 as the mesh routing protocol \cite{neumannBMX6}.

\subsection{Characterization: Bandwidth-Hungry}

In the following, we characterize the network performance of QMP network. Our goal is to determine the key features of the network and its nodes; in particular to understand the network metrics that could help us to design new heuristic frameworks for intelligent service placement in community networks \cite{AINTEC}.
Measurements have been obtained by connecting via SSH to each QMPSU OR and running basic system commands available in the QMP distribution. This method has the advantage that no changes or additional software need to be installed in the nodes. Live measurements have been taken hourly over the last 5 months, starting from October 2015 to February 2016. We use this data to analyse main aspects of QMP network. %We present the average values in the graphs.

\begin{figure*}[htb]
  \begin{minipage}[t]{0.32\linewidth}
    \centering
    \includegraphics[scale=0.5]{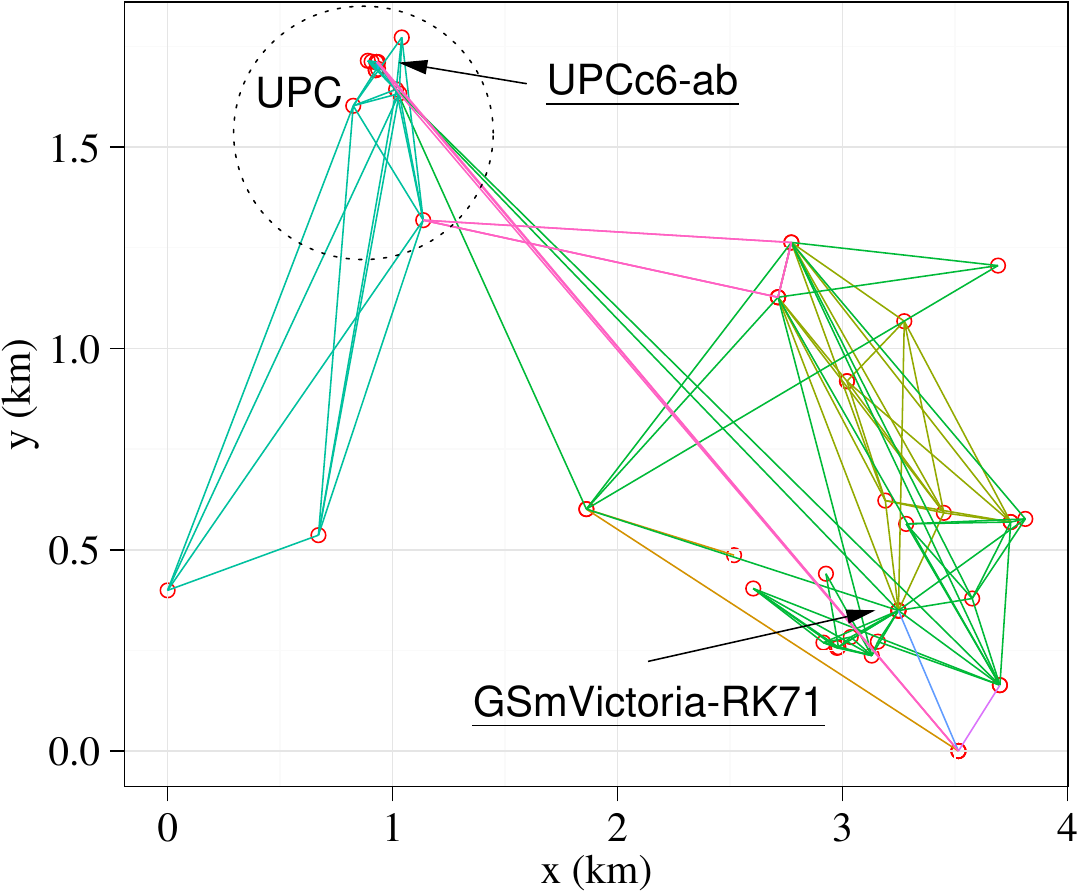}\\[-2mm] 
    \caption{QMPSU network. Two main gateways are underlined.}\label{fig:qmpsantsupc}
  \end{minipage}\hspace{0.02\linewidth}%\hfill
  \begin{minipage}[t]{0.31\linewidth}
  \centering
    \includegraphics[scale=0.5]{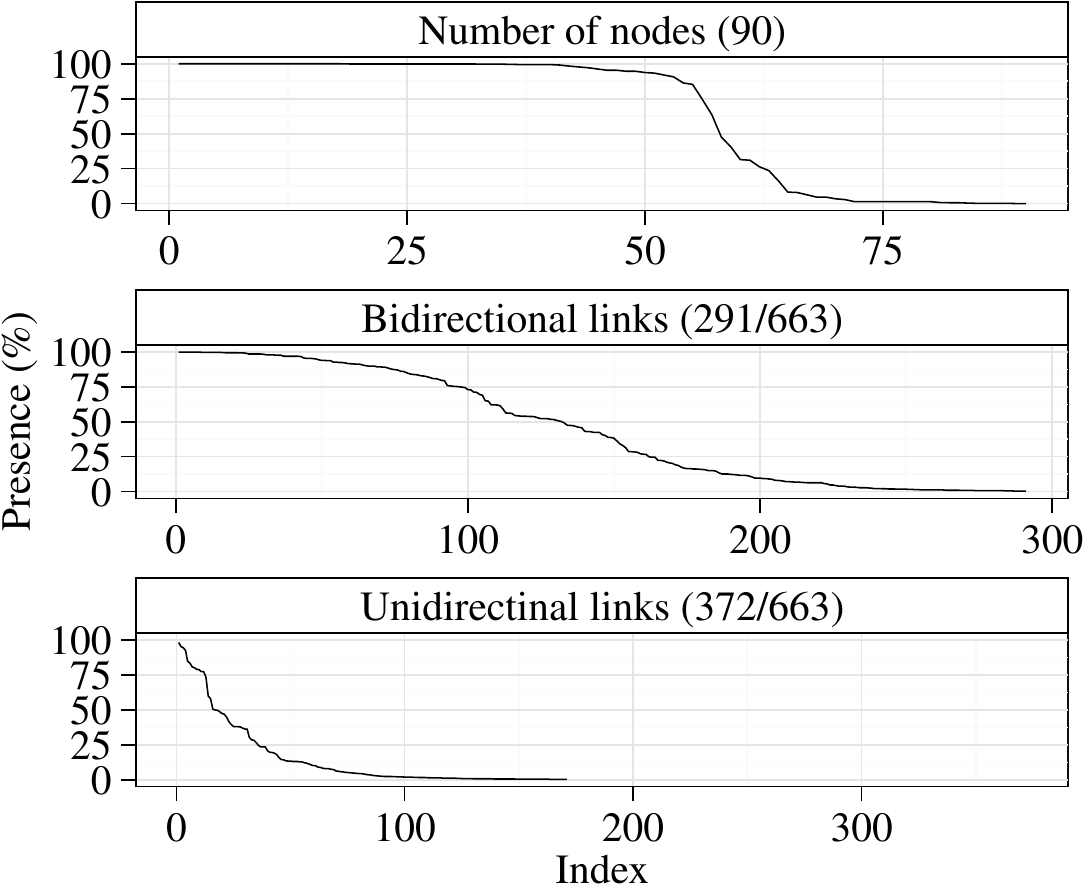}\\[-2mm] 
    \caption{Nodes and links presence.}\label{fig:presence}
  \end{minipage}\hfill
  \begin{minipage}[t]{0.33\linewidth}
    \centering
    \includegraphics[scale=0.5]{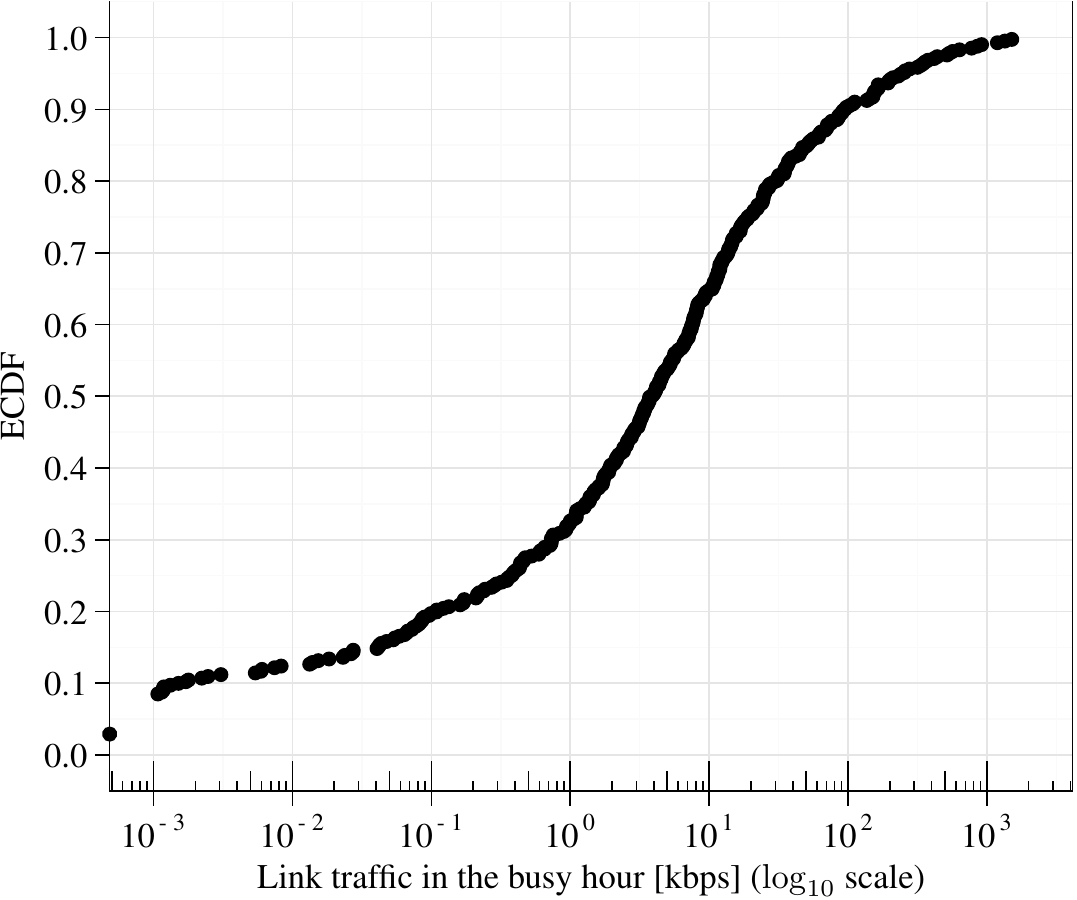}\\[-2mm] 
    \caption{Link traffic in the busy hour ECDF.}\label{fig:traffic-bh-ecdf}
  \end{minipage}\hfill
\end{figure*}

\begin{figure*}[htb]
  \begin{minipage}[t]{0.33\linewidth}
    \centering
    \includegraphics[scale=0.5]{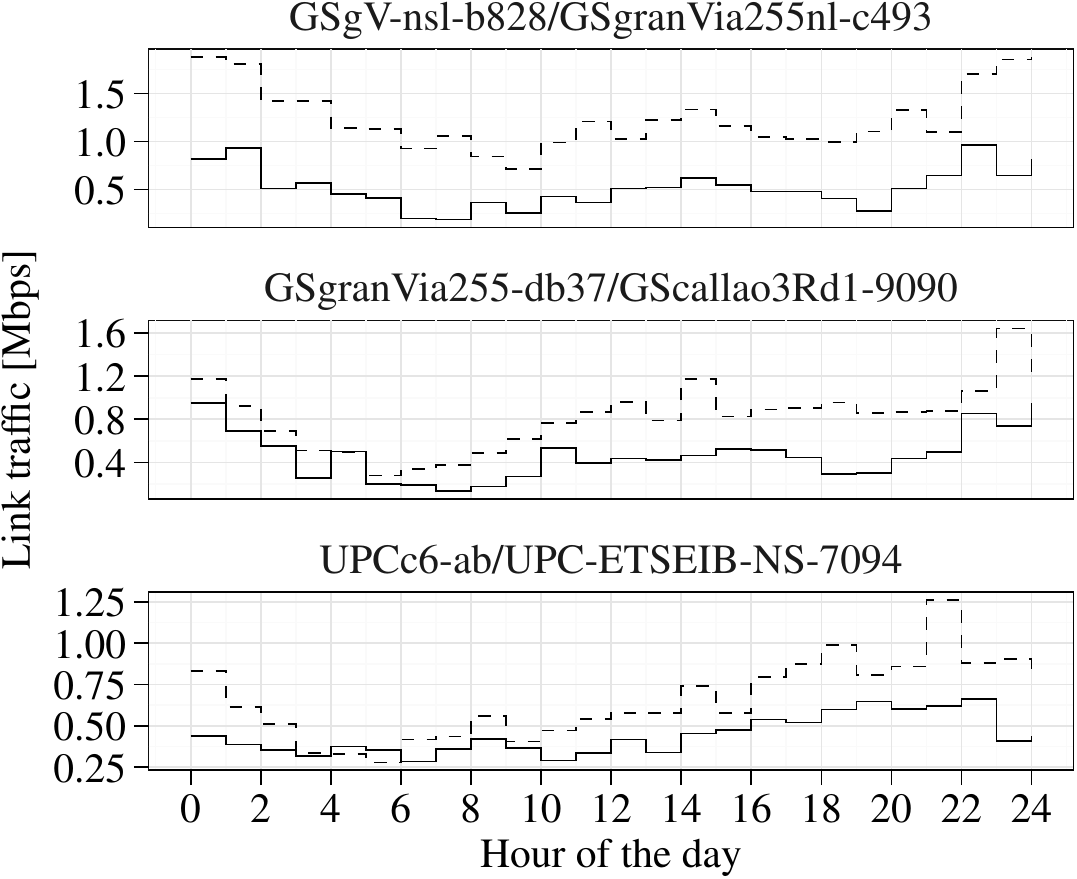}\\[-2mm] 
    \caption{Traffic in the 3 busiest links.}\label{fig:link-traffic}
  \end{minipage}\hfill
  \begin{minipage}[t]{0.32\linewidth}
    \centering
    \includegraphics[scale=0.5]{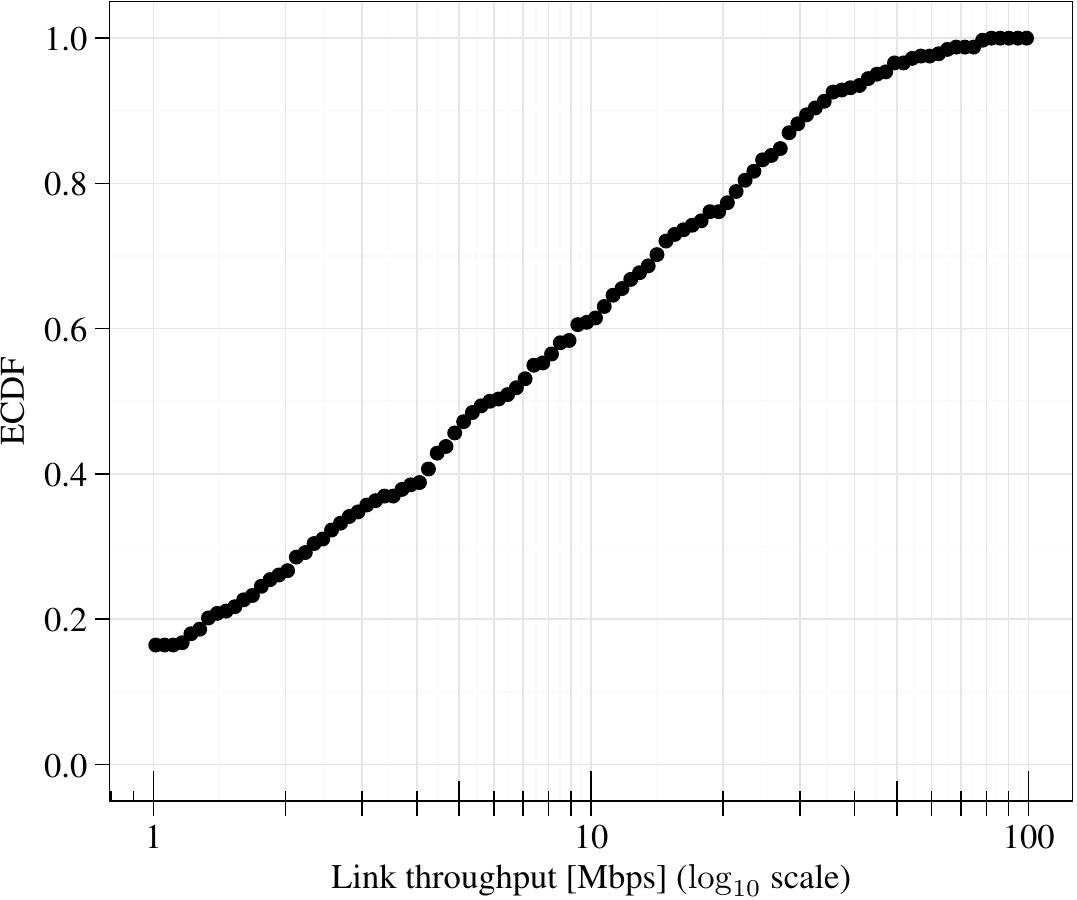}\\[-2mm] 
    \caption{Throughput ECDF.}\label{fig:bw-ecdf}
  \end{minipage}\hfill
  \begin{minipage}[t]{0.32\linewidth}
    \centering
    \includegraphics[scale=0.5]{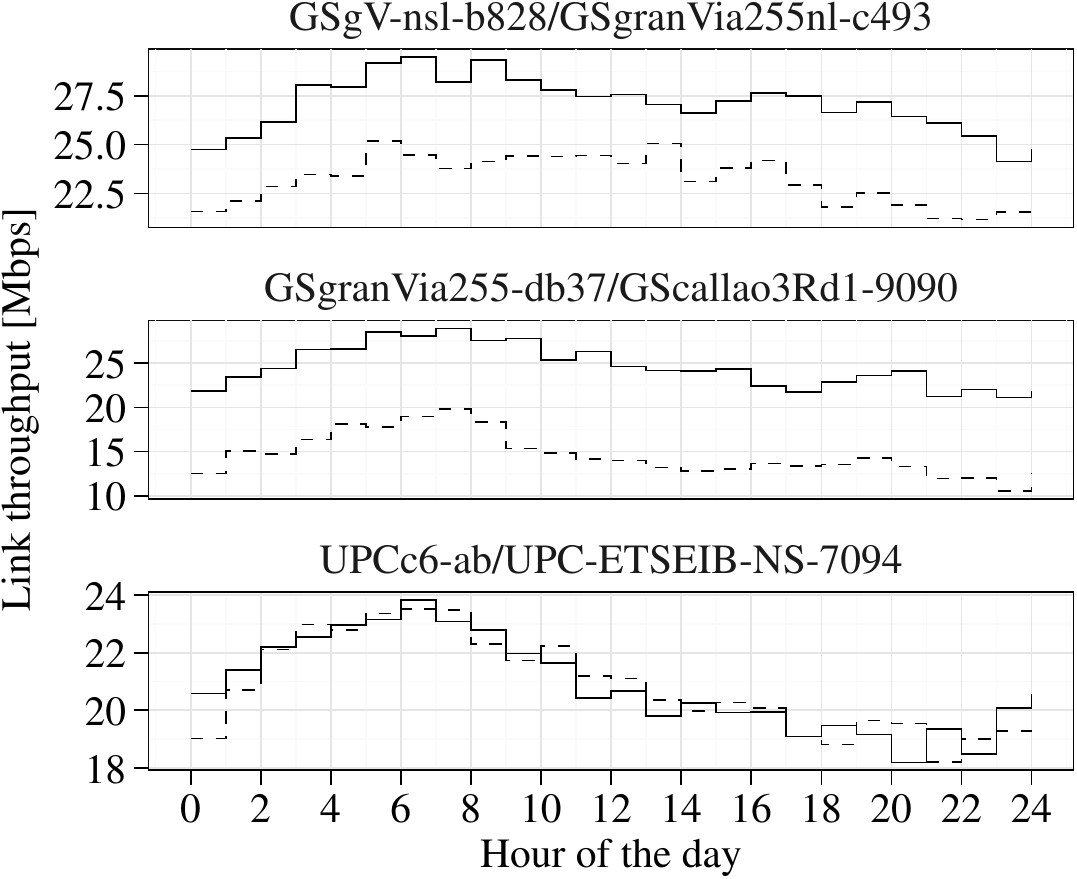}\\[-2mm] 
    \caption{Throughput in the 3 busiest links.}\label{fig:link-bw}
  \end{minipage}
\end{figure*}

Figure \ref{fig:presence} shows the node and link presence. We define presence as the percentage a given node or link is observed over the captures. Overall, 90 different nodes were detected. From those, only 61 were alive during the all measurement period, leading to a presence higher than 98\%. Around 30 nodes were missed in majority of the captures (i.e., presence less than 10\%). These are temporarily working nodes from other mesh networks and laboratory devices used for various experiments. Figure \ref{fig:presence} also reveals that 56\% of links used between nodes are unidirectional and others are bidirectional.

Figure \ref{fig:traffic-bh-ecdf}, depicts the Empirical Cumulative Distribution Function (ECDF) of the average traffic sent in each of the links in the busy hour. %\mennan{Traffic is well fitted by a mixture of 2 exponentials. Traffic that belongs to a single user and aggregate traffic from a number of users } \liang{a mixture of 3 exponential functions? maybe I am wrong. anyway it is a minor thing.} 
The overall average traffic observed is 70 kbps . Figure \ref{fig:link-traffic} shows the average traffic in both directions (upload/download) of the three busiest links.

We characterize the wireless links of the QMP network by studying their throughput. Figure \ref{fig:bw-ecdf} shows the ECDF of the throughput of the links. The figure shows that the link throughput can be fitted with an exponential distribution with mean~21.8 Mbps. In order to see the variability of the throughput, Figure \ref{fig:link-bw} shows the throughput averages in both directions (upload and download) of the three busiest links (same links as in Figure \ref{fig:link-traffic}). When we compare the Figure \ref{fig:link-bw} and Figure \ref{fig:link-traffic}, we observe that the throughput is slightly affected by the traffic in the links. Solid and dashed lines are used to identify the measurements on each direction of the links (dashed line for download, solid line for upload). It is interesting to note that the asymmetry of the throughputs measured in both directions it not always due to the asymmetry of the user traffic. For instance (node GSgranVia255), around 6am,
when the user traffic is the lowest and equal in both directions, the asymmetry of the links throughputs observed in Figure \ref{fig:traffic-bh-ecdf} remains the same. We thus conclude that this asymmetry must be due to the link characteristics, as level of interferences present at each end, or different transmission powers.

A significant amount of applications that run on Guifi.net and QMP network are network-intensive (bandwidth and delay sensitive), transferring large amounts of data between the network nodes \cite{COMNET}. The performance of such kind of applications depends not just on computational and disk resources but also on the network bandwidth between the nodes on which they are deployed. Therefore, the placement of such services in the network is of high importance. Here are some observations (features) that we captured from the measurements in QMP network:
\begin{itemize}
    \item QMP network is highly dynamic and diverse due to many reasons, e.g., its community nature in an urban area; its decentralised organic growth with extensive diversity in the technological choices for hardware, wireless media, link protocols, channels, routing protocols etc.; its mesh nature in the network etc. The current network deployment model is based on geographic singularities rather than QoS. The network is not scale-free. The topology is organic and different for conventional ISP network.  
    
    \item The resources are not uniformly distributed in the network. Wireless links are with asymmetric quality for services (30\% of the links have a deviation higher than 30\%). We observed a highly skewed traffic pattern (Figure \ref{fig:traffic-bh-ecdf}) and highly skewed bandwidth distribution (Figure \ref{fig:bw-ecdf}).  
\end{itemize}

Currently used \textit{organic (random) placement scheme} in Guifi.net community network is not sufficient to capture the dynamics of the network and therefore it fails to deliver the satisfying QoS. The strong assumption under random service placement, i.e., uniform distribution of resources, does not hold in such environments. Furthermore, 
the services deployed have different QoS requirements. Services that require intensive inter-component communication (e.g streaming service), can perform better if the replicas (service components) are placed close to each other in high capacity links \cite{CLOUDCOM15}. On other side, bandwidth-intensive services (e.g., distributed storage, video-on-demand) can perform much better if their replicas are as close as possible to their final users (e.g., overall reduction of bandwidth for service provisioning).
Our goal is to build on this insight and design a network-aware service placement algorithm that will improve the service quality and network performance by optimizing the usage of scarce resources in community networks such as bandwidth.

%\mennan{Not sure whether to consider only bandwidth or generalize it for other network metrics}.
%\liang{imo, it is better to be specific, so bandwidth-aware is good, also because NASP aims to maximise bandwidth.}

%\liang{Based on the measurement, we can derive a list of observations/features of QMP network, which we will use to derive our motivation of "smart" service placement, which will further connect to Section 3.}

%\liang{Several perspectives in my mind which might be useful for you to argue are (just suggestions, pick the one useful for you): 1) from guifi.net topological characteristics. The topology is organic and different for conventional ISP network, judged by Figure 2. 2) Usage pattern like localised communication, data sharing, rendezvous services, storage and etc, so we reduce latency and save bandwidth; 3) highly skewed traffic pattern vs. also highly skewed bandwidth distribution, the tension between these two also requires us to design smart placement algorithm. 4) node availability and how to capture the dynamics in the network and etc. Basically, pitch clearly that the real context drives us to design this algorithm. You can bring in the fact that we already started doing that by random placement, however, then you need to point out why it is not sufficient, e.g., not being able to capture the high dynamics in a typical community network so fail to deliver the satisfying QoS, then we propose our solution to address this issue.}

\label{sec:model}
\section{Bandwidth-Aware Placement}

The deployment and sharing of services in community networks is made available through \emph{community network micro-clouds} (CNMCs). The idea of CNMC is to place the cloud closer to community end-users, so users can have fast and reliable access to the service. To reach its full potential, a CNMC needs to be carefully deployed in order to utilize the available bandwidth resources.

\subsection{Assumptions}

In a CNMC, a server or low-power device is directly connected to the wireless base-station providing cloud services to users that are either within a reasonable distance or directly connected to base-station. These nodes are core-graph nodes what we call in Guifi.net. It is important to remark that the services aimed in this work are at infrastructure level (IaaS), as cloud services in current dedicated datacenters (we assume QMP nodes are core-graph nodes). Therefore the services are deployed directly over the core resources of the network (nodes in the core-graph) and accessed by base-graph clients. Services can be deployed by Guifi.net users or administrators. 

The services we consider can be centralized or distributed. The distributed services can be composite services (non-monolithic) built from simpler parts, e.g., video streaming. These parts or components of service would create an overlay and interact with each other to offer more complex services. A service may or may not be tied to a specific node of the network. Each nodes can host one or more services. 

 In this work we assume offline service placement approach where single or a set of application are placed "in one shot" in the underlying physical network. We might rearrange the placement of the same service over the time because of the service performance fluctuation (e.g. weather conditions, node availability, changes in use pattern, and etc.). We do not consider real-time service migration. 
 %%% \liang{I think we need a bit more justifications. E.g., by saying our measurement on guifi.net has shown usage follows a strong diurnal/monthly pattern and rather stable, therefore we only update the placement if there is a significant change in user pattern, or others reasons.}

%%% \mennan{We are considering community network homegrown services (inside CNs). In this work we do not consider online services e.g., online streaming, DropBox etc}
%%% \liang{What's the difference between these two? Because of proprietary code? If so, we can simply say we focus on open (source) services, otherwise, we need some business agreement between users and service providers (like Facebook).}

\subsection{Formulation and Notations}
We call the community network the \emph{underlay} to distinguish it from the \emph{overlay} network which is built by the services. The underlay network is supposed to be connected and we assume each node knows whether other nodes can be reached (i.e., next hop is known). We can model the underlay graph as: $G \gets (OR, L)$ where OR is the set of outdoor routers present in the CNs and $L$ is the set of wireless links that connects them. 

Let $f_{ij}$ be the bandwidth of the path to go from node $i$ to node $j$. We want a partition of $k$ clusters: $S \gets {S_1, S_2, S_3,...,S_k}$ of the set of nodes in the mesh network. The cluster head $i$ of cluster $S_i$ is the location of the node where the service will be deployed.The partition maximizing the bandwidth from the cluster head to the other nodes in the cluster is given by:
\begin{equation}
\label{equ1}
    \operatorname{arg\,max}_S \sum_{i=1}^k \sum_{j\in Si} f_{ij} 
\end{equation}

\label{sec:algorithm}
\subsection{Proposed Algorithm: BASP}

%\liang{I merged this section with Section 3. Otherwise, this section looks too thin by itself. You can revert to the old structure if you think that is better.} \mennan{Thanks ! I think this structure looks better. Maybe we need to change the name of section 3 now}

We designed a bandwidth-aware algorithm that allocated services taking into account the bandwidth of the network. We take a network snapshot (capture) from QMP network regarding the bandwidth of the links \footnote{http://tomir.ac.upc.edu/qmpsu/index.php?cap=56d07684}. Our bandwidth-aware service placement algorithm BASP (see Algorithm \ref{alg:basp}) runs in three phases. 

\begin{algorithm}[t]
  \caption{Bandwidth-aware Service Placement (BASP)}
  \label{alg:basp}
  \begin{algorithmic}[1]
    \Require{$G(V_n,E_n)$}\Comment{Network graph}
    \Statex $S \gets {S_1, S_2, S_3,...,S_k}$ \Comment{$k$ partition of clusters}
    \Statex $bw_i$ \Comment{bandwidth of node $i$}
    
    \item[]
    \Procedure{PerformKmeans}{$G, k$}
     \State\Return{$S$}
    \EndProcedure
     
     \Procedure{FindClusterHeads}{$S$}
        \State $clusterHeads \gets list()$
        \ForAll{$k \in S$}
        \ForAll{$i \in S_k$}
      \State $bw_i \gets 0$
      \ForAll{$j \in setdiff(S,i)$}
       \State $bw_i \gets bw+estimate.route.bandw(G,i,j)$
      \EndFor
    \State $clusterHeads \gets \max{bw_i}$
    \EndFor 
    \EndFor
    \State\Return{$clusterHeads$}
   
    \EndProcedure
    
     \Procedure{RecomputeClusters}{$clusterHeads, G$}
       \State $S\prime \gets list() $
       \ForAll{$i \in clusterHeads$}
            \State $cluster_i \gets list()$
            \ForAll{$j \in setdiff(G,i)$}
                \State $bw_j \gets estimate.route.bandw(G,j,i)$
                \If{$bw_j$ is best from other nodes $j$}
                    \State $cluster_i \gets j$
                \EndIf
            \State $S\prime \gets cluster_i$
            \EndFor
            \EndFor
        \State\Return{$S\prime$}    
    \EndProcedure
  \end{algorithmic}
\end{algorithm} 

(i) Initially, we use the naive k-means partitioning algorithm in order to group nodes based on their geo-location. The idea is to get back clusters of locations that are close to each other. The k-means algorithm forms clusters of nodes based on the Euclidean distances between them, where the distance metrics in our case are the geographical coordinates of the nodes. In traditional k-means algorithm, first, $k$ out of $n$ nodes are randomly selected as the cluster heads (centroids). Each of the remaining nodes decides its
cluster head nearest to it according to the Euclidean distance. After each of the nodes in the network is assigned to one of $k$ clusters, the centroid of each cluster is re-calculated. Grouping nodes based on geo-location is in line with how Guifi.net is organized. The nodes in Guifi.net are organized into a tree hierarchy of \emph{zones} \cite{VegaCN}. A zone can represent nodes from a neighborhood or a city. Each zone can be further divided in child zones that cover smaller geographical areas where nodes are close to each other. From the service perspective we consider placements inside a particular zone. 

(ii) The second phase of the algorithm it is based on the concept of finding the cluster head maximizing the bandwidth between the head and member nodes of the cluster, formed in the first phase of the algorithm. The cluster heads computed in this phase are the ones having the maximum bandwidth to the other nodes in the cluster $S_k$. The cluster heads are node candidates for service placement. 

(iii) The third and last phase of the algorithm includes reassigning the nodes to the selected cluster heads having the maximum bandwidth.

 %limit the choice of the cluster heads to solve (1) to be inside the sets of clusters obtained using k-means.

%To decide what is the optimum, for each choice of the cluster heads we put each of the other nodes in the cluster where fij is the highest.
%The Bw between two nodes is estimated as the Bw of the link having the minimum Bw in the shortestpath.

%\liang{Start with with some intuitive explanations of the heuristic? E.g., the purpose of Kmean ... group based on geo-location? for load balancing purpose?}

%\liang{Also need to introduce notations and formal optimisation model so that we can compare the performance gap between the optimum and heuristic. Maybe put this formulation in the model section?}
%\mennan{Done ! I moved the notations in model section}

Regarding computational complexity, the naive brute force method can be estimated by calculating the \textit{Stirling number of the second kind} \cite{Stirling} which counts the number of ways to partition a set of $n$ elements into $k$ nonempty subsets, i.e., $\frac{1}{k!} \sum_{j=0}^{k} (-1)^{j-k}  {{n}\choose{k}} j^n$ $\Rightarrow$ $\mathcal{O}(n^{k} k^n)$. However, for BASP, finding the optimal solution to the k-means clustering problem if $k$ and $d$ (the dimension) are fixed (e.g., in our case $n=54$, and $d=2$), the problem can be exactly solved in time $\mathcal{O}(n^{dk+1}\log{n})$, where n is the number of entities to be clustered. The complexity for computing the cluster heads in phase two is $\mathcal{O}(n^2)$, and $\mathcal{O}(n)$ for the reassigning the clusters in phase three. Therefore, the overall complexity of BASP is $\mathcal{O}(n^{2k+1}\log{n})$, which is significantly smaller than the brute force method.

%\liang{please double check I didn't make any stupid mistakes.} \mennan{It seems its OK !}
%\liang{Besides figures, it might be good to give a brief analysis on the complexity of the algorithm, i.e., , one short paragraph should be enough.}

%I am commenting the comments that are fixed
%\liang{Ok ... I think the naming of the algorithm seems a bit problematic. OPT is a bit confusing, it is better we come up a cute name for the proposed algorithm. For K-Mean, you can call it KM Naive ...}

\label{sec:evaluation}
\section{Preliminary Evaluation}

%%% \liang{since we compare to Random placement, we also argue NASP outperform this default strategy currently used by guifi.net, you need to give a very brief intro about how this random strategy work, and put it in the proper section, either in this section or in section 2 ... what do you think?} \mennan{already wrote something in section 2.2 but not sure whether is enough ? Well, the default deployment  model is based on geo-graphic singularities rather than network QoS}

Solving the problem stated in Equation \ref{equ1} in brute force for any number of $N$ and $k$ is NP-hard.
%\liang{not consistent notation: N and n}
%%% \liang{are you sure about this statement? Shouldn't it be "Solving the problem stated in Equation \ref{equ1} in brute force way for any number of $N$ and $k$ is NP-hard and very costly."} \mennan{Yes ! Sorry I wrote it lat night without thinking }
For this reason we came out with our heuristic. Initially we used k-means algorithm for a first selection of the clusters. Then, we limit the choice of the cluster heads to be inside the sets of clusters obtained using k-means. Inside these clusters we computed the cluster heads having the maximum bandwidth to the other nodes. To emphasise the importance of phase two and three, in this section we compare \emph{BASP} to \emph{Naive K-Means} which partitions the nodes into $k$ groups such that the sum of squares from nodes to the assigned cluster heads (centroids) is minimized. At the minimum, all cluster heads in \emph{Naive K-Means} are at the mean of their Voronoi sets (the set of nodes which are nearest to the cluster heads).
%\liang{did we ever mention how we choose cluster head in naive k-mean? need to mention how it is different from basp so that we emphasise the importance of phase 2 and 3. I added one small piece of sentence in the first paragraph, if you think it is ok, you can extend from there. basically, we need to let reviewers know what we will compare in this section as early as possible.} \mennan{Yes ! I extended your paragraph. IN fact I already wrote smth in Section 3.3, but its better to make it clear again here}

\begin{figure}[t]
\centering
\includegraphics[width=3.2in,height=1.8in]{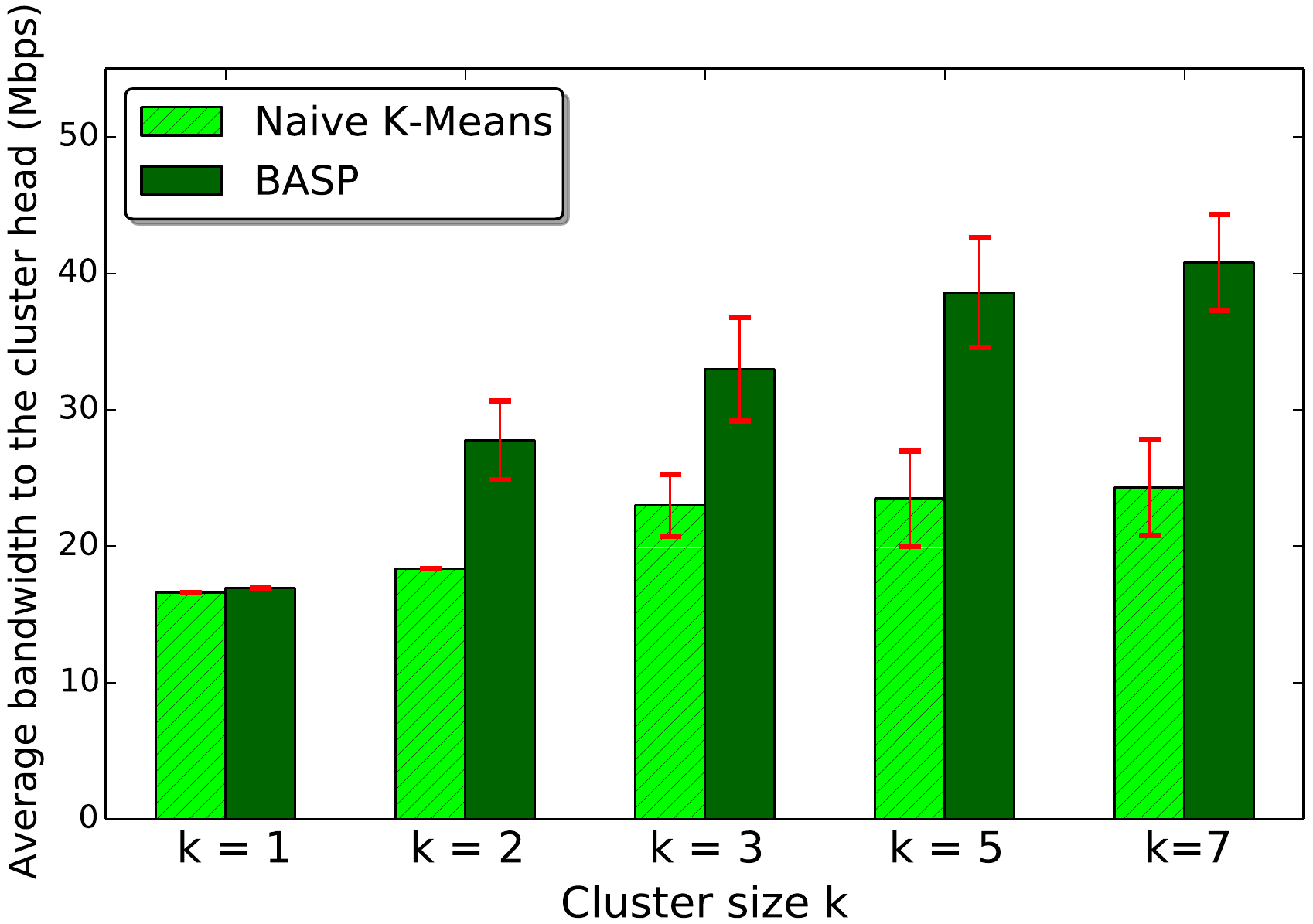}
\caption{Average bandwidth to the cluster heads}
\label{fig:bw_clusters}
\end{figure}

Our experiment is comprised of 5 runs and the presented results are averaged over all the successful runs. Each run consists of 15 repetitions. Figure \ref{fig:bw_clusters} depicts the average bandwidth to the cluster heads obtained with \emph{Naive K-Means} algorithm and our \emph{BASP} algorithm. Figure reveals that for any number of $k$, our \emph{BASP} algorithm outperforms the \emph{Naive K-Means} algorithm. For k=2 the average bandwidth to the cluster head is increased from 18.3 Mbps (obtained with naive k-means) to 27.7 Mbps (obtained with our BASP algorithm) i.e., 40\% increase. The biggest increase of 50\% is when k=7. Based on the observations from the Figure \ref{fig:bw_clusters}, the gap between two algorithms is growing as $k$ increases. K increases as network grows.  

Note that our heuristics enables us to select nodes (cluster heads) that provide much higher bandwidth then any other random or naive approach. But, if we were about to look for the optimum bandwidth within the clusters (i.e., optimum average bandwidth for the cluster), then this problem would end up to be an NP-hard. Finding the solution is NP-hard, because finding the optimum entails running our algorithm for all the combinations of size $k$ from a set of size $n$ . This is a combinatorial problem that becomes intractable even for small sizes of $k$ or $n$ (e.g., $k=5$, $n=54$). For instance, if we would like to find the optimum bandwidth for a cluster of size k=3, then the algorithm need to run for every possible (non repeating) combination of size 3 from the set of size 54. That is for 54 nodes we would end up having~25K combinations ($choose(54,3)$), or 25K possible nodes to start with. We managed to do this and the optimum average bandwidth obtained was 62.7 Mbps. The optimum bandwidth obtained for $k=2$ was 49.1 Mbps, and for $k=1$ was 16.9 Mbps. However the computation time took very long (65 hours for $k=3$, 30 minutes for $k=2$ etc.), comparing to BASP where it took 23 seconds for $k=3$ and 15 seconds for $k=2$. Table \ref{tab:centrality} shows the BASP improvement over Naive K-Means algorithm. Furthermore, Table \ref{tab:centrality} shows some centrality measures and some graph properties obtained for each cluster head. To summarize, BASP is able to achieve good bandwidth performance with very low computation complexity.

\begin{figure}[hb]
\centering
\includegraphics[width=2.8in,height=2.8in]{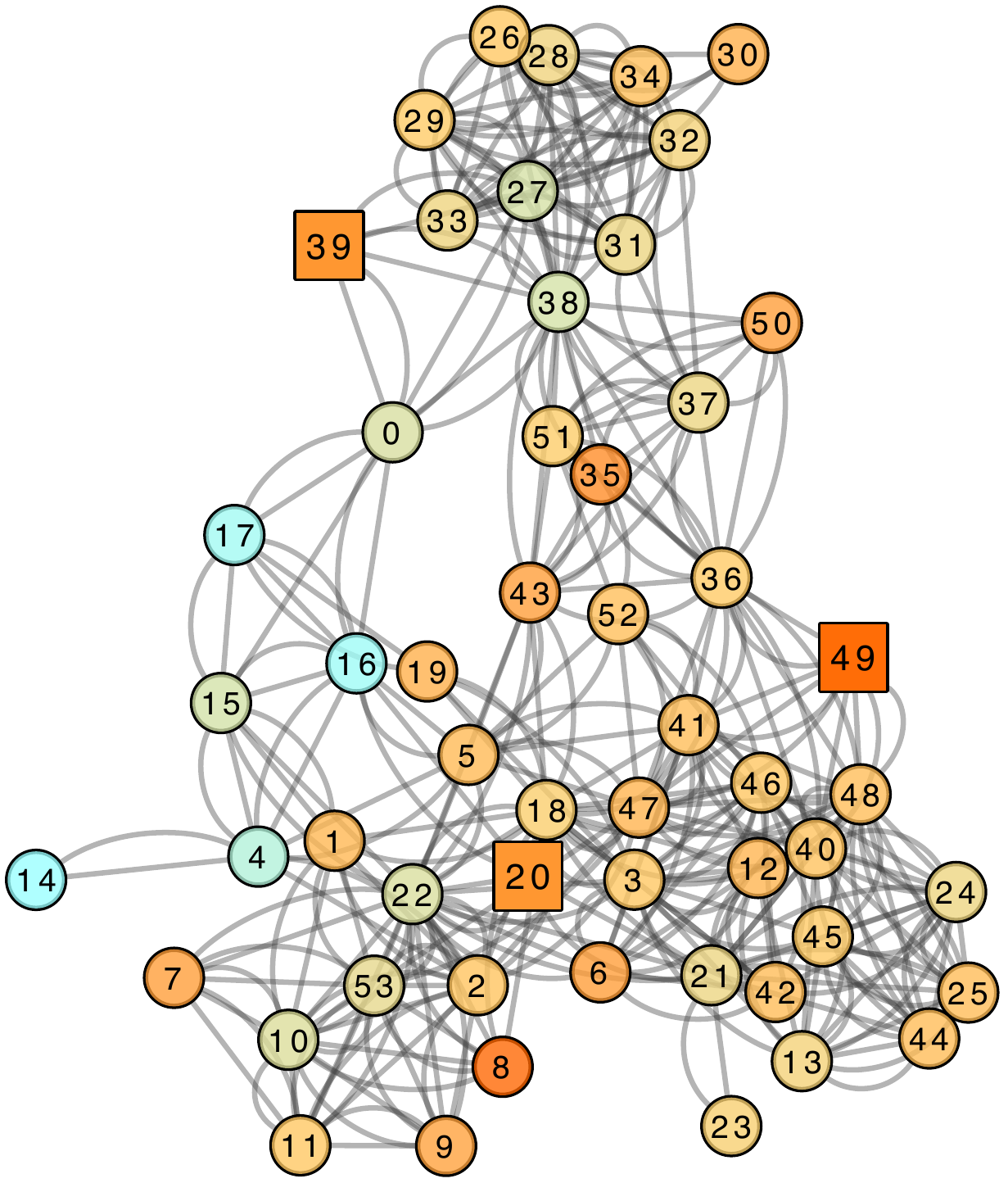}
\caption{Neighborhood Connectivity}
\label{fig:neighborhood}
\end{figure}

\begin{table*}
\caption{Centrality measures for cluster heads}
\label{tab:centrality}
\centering
\begin{adjustbox}{max width=\textwidth}
\begin{tabular}{|*{13}{c|}}
\hline
\multicolumn{1}{|c}{} & \multicolumn{1}{|c}{\textbf{k=1}} & \multicolumn{2}{|c}{\textbf{k=2}} & 
\multicolumn{3}{|c|}{\textbf{k=3}} & \multicolumn{5}{|c|}{\textbf{k=5}} \\ \hline 
 Clusters [node id] & C1 [27]  & C1 [20] &C2 [39] & C1 [20] & C2 [39]  &C3 [49] & C1 [20] &C2 [4] &C3 [49] & C4 [51] &C5 [39]  \\ \hline
 \textbf{Head degree} & 20 & 6 & 6 & 6 & 6 & 10 & 6 & 10 & 10 & 12 & 6 \\ \hline
\textbf{Neighborhood Connectivity} & 7.7 & 9.6 & 9.6 & \textbf{9.6} & \textbf{9.6} & \textbf{10.8} & 9.6 & 8.7 & 10.8 & 8.1 & 9.6  \\ \hline
\textbf{Diameter} & 6 & 5 & 3 & 4 & 3 & 5 & 4 & 2 & 3 & 1 & 3  \\ \hline
\textbf{Naive K-Means Bandwidth [Mbps]} & 16.6 & \multicolumn{2}{|c|}{18.3}& \multicolumn{3}{|c|}{23} & \multicolumn{5}{|c|}{23.4}   \\ \hline
\textbf{BASP Bandwidth [Mbps]} & 16.9 & \multicolumn{2}{|c|}{27.7}& \multicolumn{3}{|c|}{32.9} & \multicolumn{5}{|c|}{38.5}   \\ \hline
\textbf{BASP Running Time} & 7 sec & \multicolumn{2}{|c|}{15 sec}& \multicolumn{3}{|c|}{23 sec} & \multicolumn{5}{|c|}{30 sec}   \\ \hline
%\textbf{Optimum Running Time} & 1 min & \multicolumn{2}{|c|}{30 min}& \multicolumn{3}{|c|}{65 hours} & \multicolumn{5}{|c|}{400+ hours }   \\ \hline

%\textbf{Optimum} & 16.92 Mbps & \multicolumn{2}{|c|}{49.1 Mbps}& \multicolumn{3}{|c|}{62.7 Mbps} & \multicolumn{5}{|c|}{? Mbps}   \\ \hline

\end{tabular}
\end{adjustbox}
\end{table*}

\textbf{Correlation with centrality metrics:} Figure \ref{fig:neighborhood} shows the neighborhood connectivity graph of the QMP network.The neighborhood connectivity of a node $n$ is defined as the average connectivity of all neighbors of $n$. In the figure, nodes with low neighborhood connectivity values are depicted with bright colors and high values with dark colors. It is interesting to note that the nodes with the highest neighborhood connectivity are the the cluster heads obtained with our BASP algorithm. The cluster heads (for k=2 and k=3) are illustrated with a rectangle in the graph. A deeper investigation into the relationship between service placement and network topological properties is out of the scope of this paper and will be reserved as our future work.

\label{sec:related-work}
\section{Related Work}

Service placement is a key function of cloud management systems. Typically, by monitoring all the physical and virtual resources on a system, it aims to balance load through the allocation, migration and replication of tasks. 

\textbf{Data centers:} Choreo \cite{Choreo} is a measurement-based method for placing applications in the cloud infrastructures to minimize an objective function such as application completion time. Choreo makes fast measurements of cloud networks using packet trains as well as other methods, profiles application network demands using a machine-learning algorithm, and places applications using a greedy heuristic, which in practice is much more efficient than finding an optimal solution. In \cite{ambient} the authors proposed an optimal allocation solution for ambient intelligence environments using tasks replication to avoid network performance degradation. Volley \cite{VolleyNSDI2010} is a system that performs automatic data placement across geographically distributed datacenters of Microsoft. Volley analyzes the logs or requests using an iterative optimization algorithm based on data access patterns and client locations, and outputs migration recommendations back to the cloud service.

\textbf{Distributed Clouds:} There are few works that provides service placement in distributed clouds with network-aware capabilities. The work in \cite{SIGCOMM12} proposes efficient algorithms for the placement of services in distributed cloud environment. Their algorithms need input on the status of the network, computational resources and data resources which are matched to application requirements. In \cite{www12} authors propose a selection algorithm to allocate resources for service-oriented applications and the work in \cite{INFOCOM12} focuses on resource allocation in distributed small datacenters. 

\textbf{Service Migration:} Regarding the service migration in distributed clouds, few works came out recently. The authors in \cite{ServiceMigration1} and \cite{ServiceMigration2} study the dynamic service migration problem in mobile edge-clouds that host cloud-based services at the network edge. They formulate a sequential decision making problem for service migration using the framework of Markov Decision Process (MDP) and illustrate the effectiveness of their approach by simulation using real-world mobility traces of taxis in San Francisco. The work in \cite{ServiceMigration3} studies when services should be migrated in response to user mobility and demand variation.

%From the review of the related work it can be seen that the reviewed studies are not conducted in the context of the community networks, which we address as scenario. 

While our focus in this paper is to design a low-complexity service placement heuristic for community network clouds to maximise bandwidth, another closely related work is \cite{davide} which proposed several algorithms that minimize the coordination and overlay cost along a network.

\label{sec:conclusion}
\section{Conclusion}
In this paper, we first motivated the need for bandwidth-aware service placement on community network micro-cloud infrastructures. Community networks provide a perfect scenario to deploy and use community services in contributory manner. Much previous work done in CNs has focused on better ways to design the network to avoid hot spots and bottlenecks.
As services become more network-intensive, they can become bottle-necked by the network, even in well-provisioned clouds. The case in community network clouds is even more hair-raising, with limited capacity of nodes and links and an unpredictable network performance. Without a network aware system for placing services, poor paths can be chosen while faster, more reliable paths go unused. 

Furthermore, we proposed a low-complexity service placement heuristic called BASP to maximise the bandwidth allocation in deploying a CNMC. We presented algorithmic details, analysed its complexity, and carefully evaluated its performance with realistic settings. Our preliminary results show that BASP consistently outperforms the currently adopted random placement in Guifi.net by 35\%. Moreover, as the number of services increases, the gain tends to increase accordingly.
%We employed a topological snapshot from Guifi.net to identify node traits in the optimal service placements. We used the data to find how network-aware metrics are important in the node selection.

As a future work, we plan to deploy our service placement algorithm in a real network segment of Guifi.net, using real services and quantify the performance and effects of the algorithm.

%%% \liang{it is better to also acknowledge some shortcomings and future directions.}

%ACKNOWLEDGMENTS are optional
%\section{Acknowledgments}
%This work was supported by several projects: the European Commission FP7 FIRE CONFINE (FP7-288535), CLOMMUNITY (FP7-317879), H2020 RIFE (H2020-644663), and the Spanish government (TIN2013-47245-C2-1-R).

\balancecolumns
% That's all folks!
\end{document}